# Quantum wavefunction reconstruction by free-electron spectral shearing interferometry

Zhaopin Chen[1,2], Bin Zhang[3], Yiming Pan[1*], Michael Krüger[1,2]

1. Department of Physics, Technion – Israel Institute of Technology, Haifa 3200003, Israel
2. Solid State Institute, Technion – Israel Institute of Technology, Haifa 3200003, Israel
3. Department of Electrical Engineering Physical Electronics, Tel Aviv University, Ramat Aviv 6997801, Israel

**Abstract**

We propose a novel spectral method for reconstructing quantum wavefunction of an electron pulse, free-electron spectral shearing interferometry (FESSI). We employ a Wien filter to generate two time-delayed replicas of the electron wavepacket and then shift one replica in energy using a light-electron modulator driven by a mid-infrared laser. As a direct demonstration, we numerically reconstruct an ultrashort electron pulse with a kinetic energy of 10 keV. FESSI is experimentally feasible and enables us to fully determine distinct orders of spectral phases and their physical implications, providing a universal approach to characterize ultrashort electron pulses.

A free-space single electron wavepacket perfectly unites in itself the wave-particle duality of quantum mechanics [1]. At a given point in space, an electron wavepacket is characterized by a spectral amplitude and phase, much like the electric field of a short light pulse. A Fourier transformation of both quantities is then able to reveal the full quantum wavefunction of the electron wavepacket. While it is straightforward to measure the spectrum with a precise electron spectrometer, accessing the spectral phase of electron wavepackets in free space remains challenging.

Recent research demonstrates that ultrafast laser pulses can generate and manipulate the wavefunction of free electrons with great coherence and accuracy [2–4]. For instance, electrons generated from laser-excited photocathodes [5] and nanoscale metallic tips [4,6–8] can produce femtosecond electron pulses for a wide range of applications in ultrafast microscopy and diffraction. Control of the photoemission process is possible down to the attosecond time scale [9,10]. Nanotip sources in particular enable high spatial and temporal coherence [11–14]. Manipulation of free electrons with light is also advancing rapidly, for example, using photon-induced nearfield electron microscopy (PINEM) [12,15–18], the Kapitza-Dirac effect [19], attosecond electron streaking [20–22] and dielectric laser acceleration (DLA) [23–25]. Such manipulation can further reduce the electron pulse duration [21] or lead to the formation of pulse trains [12,26,27]. Detection methods for determining the duration of femtosecond electron pulses include ponderomotive scattering [28], streaking with infrared and THz fields [20,21,29], or spectral quantum interference for the regularized reconstruction of free-electron states (SQUIRRELS [26]). However, in these systems, the electron is either seen as an ensemble of point particles (scattering and streaking methods) or plane waves (SQUIRRELS). Up to now, tracing the quantum phase of a free electron wavepacket has remained elusive.

In order to construct the full temporal wavefunction of an electron, the amplitude and phase profiles must be measured, i.e., $\psi(t) = \sqrt{\rho(t)}e^{i\varphi(t)}$. Using attosecond streaking [20], the longitudinal density distribution $\rho(t)$ of an electron pulse may be determined. However, the quantum phase $\varphi(t)$ is lost while measuring the amplitude.

Here, we propose spectral shearing interferometry for reconstructing the free electron wavefunction. The most basic element of spectral shearing interferometry is a spectrometer, which records the electron energy spectrum. With a resolution of 10 meV or better, commercial high-resolution electron spectrometers may directly detect the electron spectrum. Therefore, measuring the spectral phase profile is our primary concern. Specifically, we focus on the ultrafast electron wavefunction produced from a laser-excited photoelectron gun or from laser-driven modulation of electron pulses. Our approach, which we name free electron spectral shearing interferometry (FESSI), involves two key ingredients: a time delay by a Wien filter element and an energy shift by a light-electron modulator. We will discuss the principle and algorithm for FESSI in the following sections. The fundamental origin of the spectral phase is the photoelectron gun configuration and subsequent changes of the spectral phase profile changes due to free-space propagation, particle acceleration and laser-driven electron pulse modulation. By utilizing FESSI, we can calibrate and reconstruct femtosecond and attosecond electron pulses in a coherent manner. We believe our interferometry might be widely employed for coherent phase control of quantum wavefunction, characterization of ultrafast electron dynamics, and investigation of quantum foundations and applications.

Instead of directly measuring the temporal wavefunction, we would like to measure the spectral profile in the energy domain

$$\psi(E) = \sqrt{\rho(E)} e^{i\phi(E)}, \qquad (1)$$

where $\rho(E) = |\psi(E)|^2$ is the spectral density distribution measured by an electron spectrometer and $\phi(E)$ is the spectral phase. The temporal wavefunction can be obtained from the Fourier transform, $\psi(t) = \int dE\, \psi(E)\, e^{-iEt/\hbar}$. Now, the primary challenge is measuring the phase $\phi(E)$. Inspired by the optical spectral shearing interferometry for measuring the phase profile of ultrafast laser pulses, in particular spectral phase interferometry for direct electric-field reconstruction (SPIDER) [30,31], we propose an experimental setup for the direct reconstruction of the quantum phase $\phi(E)$ of ultrafast electrons, as shown in Fig. 1(a).

First, we generate a single electron pulse using a laser-excited photoelectron gun. A high-voltage electron optics system accelerates the electron wavefunction to 10 keV, sending it to an (optional) laser-driven modulator where the electron pulse can be compressed further. Then the beam encounters an electron biprism, which splits the electron wavepacket into two replicas [32]. Then, a Wien filter creates a time delay between two replicas in order to generate spectral interference fringes [33]. Furthermore, to imprint the spectral phase information onto the fringes, a light-electron modulator (LEM) accelerates one replica to produce a small energy shift with respect to the other replica. We will introduce the LEM device in the next section. After realizing the time delay between the replicas and the energy shift for one of them, we require one more electron biprism to merge two paths back into one so that the phase information is converted into the spectral intensity. Lastly, we detect the spectral fringes using a spectrometer and determine $\phi(E)$ using our reconstruction algorithm (see Fig. 2).

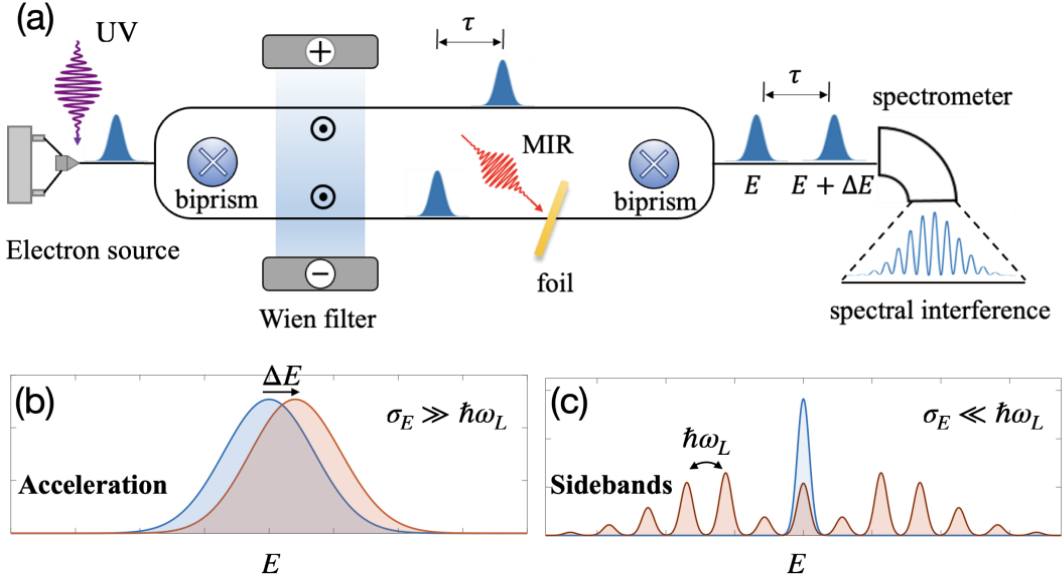

FIG. 1. (a) Setup for measurements of ultrashort electron wavepackets using free-electron spectral shearing interferometry (FESSI). An electron pulse generated by a laser-driven electron source is accelerated and split into two replicas. A Wien filter induces a time delay $\tau$ and a mid-infrared (MIR) laser pulse irradiating a thin foil shifts the energy of one replica by a small amount, leading to spectral shearing. A spectrometer measures the spectral interference. (b) In the regime of the acceleration process ($\sigma_E \gg \hbar\omega_L$), the laser-irradiated foil imparts a uniform energy shear of the electron wavepacket. (c) In the regime of the photon-induced near-field electron microscopy (PINEM) process ($\sigma_E \ll \hbar\omega_L$), sidebands will appear in the electron spectrum after interaction with the near-field at the foil.

The two key ingredients for achieving the spectral phase reconstruction are the time delay $\tau$ and a small energy shift $\Delta E$. From the FESSI setup, we obtain the spectral interference signal (see Section 2 in Supplemental Material) as

$$\phi_{\text{FESSI}}(E) = \phi(E) - \phi(E - \Delta E) - \frac{E\tau}{\hbar} \approx \Delta E \frac{\partial \phi}{\partial E} - \frac{E\tau}{\hbar}. \qquad (2)$$

Consequently, the spectral phase can be deduced from the integral $\phi(E) = \phi_0 + \frac{1}{\Delta E} \int \frac{dE'}{\hbar} \left( \phi_{\text{FESSI}} + \frac{E'\tau}{\hbar} \right)$, with an arbitrary constant phase $\phi_0$. Notice that the amplitude is directly measured by a spectrometer, allowing both the spectral profile (Eq. 1) and the temporal wavefunction of the electron pulse to be reconstructed.

To create an accurate time delay, we employ a Wien filter. A Wien filter is a device consisting of a static electric field and a magnetic field that are both perpendicular to each other and to the beam path [33] (see Fig. 1(a)), which has been extensively employed to investigate the fundamentals of quantum mechanics [32]. However, rather than longitudinal coherence, our spectral interferometry requires spectral coherence, relating to the spectral width of the electron pulse. We can observe the spectral phase signal by using a Wien filter to create interference fringes within the spectral coherence region. This indicates the shorter the pulse duration, the higher the spectral coherence. It also suggests that the spectral interferometry method is ideal for measuring ultrashort pulses.

On the other hand, in order to achieve an energy shift with high accuracy and coherence, we developed a method to precisely regulate the energy shift through the interaction with a mid-infrared (MIR) laser beam, which we refer to as light electron modulator (LEM). This LEM device is capable of generating a spectrum with net acceleration for a particle-like electron (see Fig. 1(b)) and a spectrum with energy-domain sidebands for a wave-like electron (see Fig. 1(c)) [34]. We need the LEM device to operate in the acceleration regime to obtain a net energy shift that can accelerate the electron wavefunction without affecting its spectral distribution. In our proposal, the LEM device is realized as a 50nm-thick amorphous $Si_3N_4$ foil that is illuminated by a mid-infrared laser pulse with a center wavelength of 10.33μm [22]. The laser pulse is synchronized with the arrival of the electron wavepacket at the foil. After passing through the optical field on the foil, the electron pulse is accelerated. In other words, both the spectral amplitude and phase of the electron will acquire an energy shift $\Delta E$ and a corresponding phase shift, respectively (see Fig. 1(b)). Through our LEM device, we are able to generate a coherent spectral shearing for the reconstruction.

Next, we demonstrate the mechanism of electron wavepacket acceleration by the LEM device. We start by considering the electron to be a monochromatic plane wave $\psi_0 = Ae^{-\frac{iE_0 t}{\hbar}}$ with the center energy $E_0$ and the amplitude $A$. The light-electron interaction in the LEM device results in an effective phase modulation of the initial wavefunction [26]; thus, the final electron wavefunction is given by

$$\psi_f(t) = A e^{-\frac{iE_0 t}{\hbar}} \left( \sum_n J_n(2|g|) e^{-in\omega_L t} \right), \quad (3)$$

in which $2|g| = \frac{e}{\hbar \omega_L} \int_{-\frac{L}{2}}^{\frac{L}{2}} F(z) \exp\left(-\frac{i\omega_L z}{v_0}\right) dz$ is the effective photon number exchange with $F(z)$ the electric acceleration gradient, $\omega_L$ the laser frequency, $L$ the interaction length, and $v_0$ is the central velocity of the electron. Recalling the generating function of Bessel functions, we can rewrite Eq. 3 as $\psi_f(t) = \int \frac{dE}{\hbar} [A \sum_n J_n(2|g|) \delta(E - E_0 - n\hbar\omega_L)] e^{-\frac{iEt}{\hbar}}$. Then the resulting modulated electron energy spectrum reads

$$\psi_f(E) = A \sum_n J_n(2|g|) \delta(E - E_0 - n\hbar\omega_L). \quad (4)$$

This corresponds to a typical PINEM spectrum as observed in ultrafast TEM [12,15]. By extending the plane wave electron to a pulsed electron for our purpose, we end up with a final spectrum of the form

$$\psi_f(E) = \frac{A}{(2\pi\sigma_E^2)^{\frac{1}{4}}} \sum_n J_n(2|g|) \exp\left\{-\frac{(E - E_0 - n\hbar\omega_L)^2}{4\sigma_E^2}\right\}. \quad (6)$$

Each delta-like sideband is replaced by the Gaussian sideband with a finite spectral width $2\sigma_E$. Thus, the initial electron pulse duration is given by $2\sigma_t = \hbar/\sigma_E$. Under the condition of $2\sigma_E > \hbar\Omega_L$, these sidebands overlap, and the interference between sidebands would significantly affect the final energy spectrum. To see this explicitly, we assume $|g|$ is small, and apply the Taylor expansion to these sidebands:

$$\begin{aligned}
\psi_f(E) &= \frac{A}{(2\pi\sigma_E^2)^{\frac{1}{4}}} \sum_n J_n(2|g|) \left( e^{-n\hbar\omega_L \frac{\partial}{\partial E}} \exp\left\{-\frac{(E - E_0)^2}{4\sigma_E^2}\right\} \right) \\
&= \sum_n J_n(2|g|) e^{-n\hbar\omega_L \frac{\partial}{\partial E}} \psi_0(E) \approx \sum_n J_n(2|g|) \left(1 - n\hbar\omega_L \frac{\partial}{\partial E}\right) \psi_0(E) \quad (7) \\
&= \left(1 - 2|g|\hbar\omega_L \frac{\partial}{\partial E}\right) \psi_0(E) \approx e^{-2|g|\hbar\omega_L \frac{\partial}{\partial E}} \psi_0(E) = \psi_0(E - 2|g|\hbar\omega_L),
\end{aligned}$$

where the initial spectral profile is $\psi_0(E) = \frac{A}{(2\pi\sigma_E^2)^{\frac{1}{4}}} \exp\left\{-\frac{(E-E_0)^2}{4\sigma_E^2}\right\}$ and the net energy shift is $\Delta E = 2|g|\hbar\omega_L$. In the derivation, we use the Taylor expansion formula: $f(x+\varepsilon) = e^{\varepsilon\frac{\partial}{\partial x}}f(x) \approx f(x) + \varepsilon f'(x)$, with $\varepsilon$ being a small quantity, and the following relations for Bessel functions: $J_{n-1}(z) + J_{n+1}(z) = \frac{2n}{z}J_n(z)$ and $\sum_n J_n(z) = 1$. Therefore, under the conditions of $\Delta E \ll \sigma_E$ and $\hbar\omega_L \ll \sigma_E$, the LEM device can realize spectral shearing for an electron. After passing through the LEM device, the initial electron energy spectrum given by Eq. 1 obtains a spectral shear: $\psi_f(E) = \psi_0(E - \Delta E)e^{i\phi(E-\Delta E)}$, as required for FESSI. Notably, the initial spectral amplitude $\psi_0(E)$ is not required to be Gaussian. It can be an arbitrary function with a considerably wide bandwidth. Since it is cumbersome to analytically express a general electron wavefunction modulation, we use the time-dependent Schrödinger equation (TDSE) to simulate this LEM process (see Section 1 in Supplemental Material).

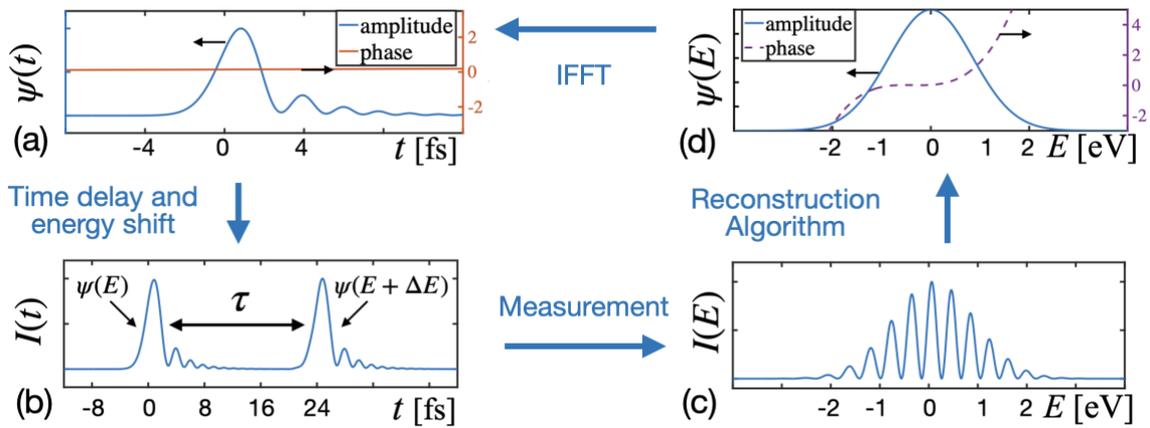

FIG. 2. Free electron wavefunction reconstruction by FESSI. (a) Temporal amplitude (blue) and phase (red) of an electron wavepacket. (b) Two replicas of the wavefunction after time shift and energy shear. (c) Spectral interference of the wavefunction replicas. (d) Spectral amplitude (blue) and the reconstructed spectral phase (purple).

Figure 2 illustrates the method of electron wavefunction reconstruction in FESSI. In Fig. 2(a), we assume a pulsed electron wavefunction with a particular amplitude and phase that is awaiting measurement. The reconstruction employs a spectral interferogram generated by the interference of two replicas of the pulse. The time delay between them is created by the Wien filter. The two replicas are identical except that their energy is displaced relative to one another (Fig. 2(b)). The resulting interferogram (Fig. 2(c)), which is the spectral intensity, is given by

$$I(E) = |\psi(E)|^2 + |\psi(E - \Delta E)|^2 + 2|\psi(E)||\psi(E - \Delta E)| \\ \times \cos(\phi(E) - \phi(E - \Delta E) - E\tau/\hbar), \tag{8}$$

where $\psi(E)$ is the spectral profile, $\Delta E$ is the spectral shear, and $\tau$ is the time delay. The first two terms are the spectra of the initial pulse and its spectral-shifted replica. The third term of Eq. 8 presents the spectral interference. By utilizing a reconstruction algorithm (for details see Fig. S1 in the Supplemental Material), we can retrieve the spectral phase profile (Fig. 2(d)). To this end, both the spectral amplitude and phase are measured, so that we can perform the inverse Fourier transform to the spectral profile (Eq. 1) and reconstruct the full temporal wavefunction. It is important to notice that the reconstructed wavefunction is located at the position of the LEM device since the spectral shearing occurs here.

The detection of the spectral phase profile is crucial since it determines the reconstruction of the electron wavefunction. In this section, we classify the spectral phase to comprehend the underlying physics that might alter or influence the spectral phase. We start with a Taylor expansion of the spectral phase, $\phi_E(E) = \phi_0 + \phi_1(E - E_0) + \frac{1}{2}\phi_2(E - E_0)^2 + \frac{1}{6}\phi_3(E - E_0)^3 + \cdots$, where $\phi_n = \partial^n \phi_E/\partial E^n$ at $E = E_0$. The zero-order constant term is only a global phase. The linear phase term $\phi_1$ relates to the arbitrary choice of the initial time $t_0$ for the electron, which is easily removed by resetting the initial time. Consequently, only higher-order phases ($n \geq 2$) dominate the reconstruction of an electron. The second-order phase $\phi_2$ results in pulse chirping in the time domain. Indeed, it is physically relevant to consider a second-order phase because a propagating electron can accumulate $\phi_2$ due to its nonrelativistic dispersion. The third-order phase $\phi_3$ is also attractive because it can be induced by a high-voltage bias when the emitted electron is pre-accelerated to the final kinetic energy, such as 10 keV. This acceleration process would introduce a third-order phase to the measured electron pulse [35,36]. Our scheme can also provide the fourth and higher order phases, although they are of minor importance in our case. For reconstructing a wavefunction, the phases of the second and third orders are most relevant. In our simulation, we generate an exemplary electron wavefunction with a Gaussian envelope function and a spectral phase that combines the second and third orders, $\phi(E) = 0.34(E - E_0)^2 + 1.05(E - E_0)^3$. Figure 3(a) shows the wavefunction side-by-side with the energy-shifted wavefunction. Here, the energy shift $\Delta E = 0.1$ eV is 11.8% of the spectral width of the pulse $2\sigma_E = 0.85$ eV.

Two constraints limit the range of time delay introduced by the Wien filter: (i) In order to observe the spectral inference within the spectral width, the delay has a lower limit, $\tau > \pi\hbar/\sigma_E$ (or, $\frac{2\pi\hbar}{\tau} < 2\sigma_E$,). (ii) Limited by the resolution $\delta_E$ of the spectrometer, the delay cannot exceed an upper limit, $\tau < 2\pi\hbar/\delta_E$. Here we assume a resolution of $\delta_E = 10$ meV such that the possible time delay ranges from 5 fs to 400 fs. We choose the delay as $\tau = 30$ fs. Figure 3(b) shows the numerical result of FESSI which demonstrates an excellent match between the reconstructed and the original spectral phases.

By a similar token, two constraints restrict the range of the energy shear: (i) We extract the phase from the a.c. term (Eq. S4 in the Supplemental Material). The ratio of intensity between the a.c. and d.c. term is approximately $\|D^{ac}/D^{dc}\| \propto \exp(-\Delta E^2/4\sigma_E^2)$. To acquire an adequate a.c. signal we must guarantee $\Delta E < 2\sigma_E$. (ii) Conversely, the energy shear cannot be too small. Otherwise, the phase contribution of ($\Delta E \frac{\partial \phi}{\partial E}$) in Eq. 2 is insignificant. In our simulation, we use $\Delta E = 0.1$ eV. Such a small energy shift for the electron pulse necessitates stable and coherent laser control of the electron acceleration, which is within the capabilities of the LEM device.

For FESSI to work, the measured electron pulse must be local in time for the LEM device to obtain a net energy shift. In analyzing the limits of the time delay and energy shear, we notice that $\sigma_E$ should be sufficiently large to observe spectral interference. This requirement is consistent with the locality condition of the electron with an intrinsic duration of $\sigma_{t0} = \hbar/2\sigma_E$ that is temporally small. However, the second and higher order spectral phases would broaden the pulse even though the intrinsic duration is small. We provide a criterion $\sigma_t(\phi) < T/4$, namely, the pulse duration should smaller than a quarter of the optical cycle $T$ (see Section 3 in SM), so that there is still a possibility of achieving spectral shear using the LEMs [37]. Due to the temporal locality of the electron wavefunction, the spectral phase variation cannot be too large. As a further test of FESSI, we examine pulse broadening in terms of different order spectral phases, including oscillatory phases (see Section 3 in Supplemental Material).

FESSI is strongly sensitive to jittering effects of different kinds. The contribution of energy jitter is negligible because the laser-electron modulator can control it coherently. However, the time jitter $\delta\tau$ is critical since it will lead to a large phase jitter, $\delta\phi \sim E_0 \delta\tau$. This phase jitter would lead to a strong blur of the interference fringes, rendering the spectral interferometry ineffective. However, state-of-the-art Wien filters can minimize time jitter to a negligible amplitude when using highly stable and precise power supplies [33]. In order to quantify the quality of spectral phase reconstruction, we define a fidelity $F = \frac{\sum |\phi_o(E)|^2}{\sum |\phi_r(E) - \phi_o(E)|^2 + \sum |\phi_o(E)|^2}$, with the original phase $\phi_o$ and reconstructed phase $\phi_r$. To investigate the potential of experimentally reconstructing the signal, we assume a tiny time jittering of about 0.001% delay time in our simulation (Fig. 3), where the fidelity $F$ is near 1. Increasing the jitter to 0.007%, the spectral interference is noticeably weaker but we still find that F = 98.1%, indicating the robustness of our method (see Fig. S2 in Supplemental Material).

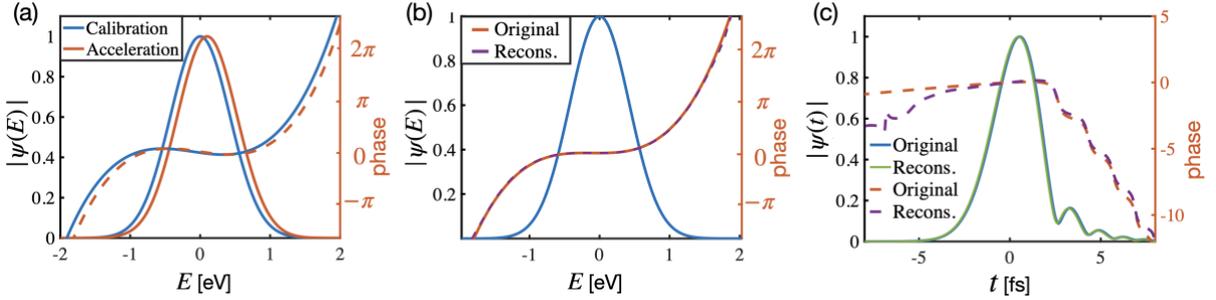

FIG. 3. Original and reconstructed free electron wavepacket with both amplitude and phase. (a) shows the original electron pulse (blue) with a spectral width of 0.85 eV and with 2nd and 3rd order dispersion. The orange curve displays the accelerated replica with an energy shift of $\Delta E = 0.1$ eV. (b) shows the spectral amplitude and the original and reconstructed phases of the electron wavefunction. With the consideration of time jitter around 0.001% of the time delay, the fidelity value for the reconstruction is $F = 99.9\%$. (c) Comparison of the original wavefunction and the reconstructed one in time.

Finally, it is worth mentioning our FESSI setup possesses the exciting capability of monitoring an ultrashort attosecond electron pulse. In Fig. S6 of the Supplemental Material, we show the successful wavefunction reconstruction of such a pulse, demonstrating that our FESSI can also address the attosecond regime.

In conclusion, we have proposed a spectral interferometry approach to characterize and reconstruct an ultrashort electron wavepacket. The principle, setup and reconstruction approach of the free-electron spectral shearing interferometry (FESSI) enable the measurement of typical electron wavepackets in ultrafast science. Since FESSI foots on existing techniques in electron microscopy and spectroscopy, we believe its experimental realization is feasible. We expect applications of FESSI as a characterization approach for femtosecond and attosecond electron pulses in the photoelectron guns, ultrafast transmission and scanning electron microscopy,

DLA, X-ray free electron lasers and strong-field physics. These domains will strongly benefit from full detection and manipulation of ultrashort electrons. Second, free electron wavefunction reconstruction can test the foundations of quantum mechanics, for example through weak measurements [38,39] and quantum tomography [40], which may verify many issues, such as the wavefunction collapse and the measurement problem [41,42], and the dispute between $\psi$-ontic and $\psi$-epistemic [43–45].

**Acknowledgments**

We thank Peter Baum, Peter Hommelhoff, Rafal Dunin-Borkowski, Kangpeng Wang and Senlin Huang for insightful discussions. This project has received funding from the European Union's Horizon 2020 research and innovation program under grant agreement No 853393-ERC-ATTIDA. We also acknowledge the Helen Diller Quantum Center at the Technion for their support.

\* Corresponding author:  yiming.pan@campus.technion.ac.il

# Supplemental Material
# Quantum wavefunction reconstruction by free-electron spectral shearing interferometry

Zhaopin Chen[1,2], Bin Zhang[3], Yiming Pan[1*], Michael Krüger[1,2]

1. Department of Physics, Technion – Israel Institute of Technology, Haifa 3200003, Israel
2. Solid State Institute, Technion – Israel Institute of Technology, Haifa 3200003, Israel
3. Department of Electrical Engineering Physical Electronics, Tel Aviv University, Ramat Aviv 6997801, Israel

1. **Schrödinger Equation solution of the light electron modulator (LEM)**

The wavepacket acceleration process can be represented by a relativistically modified Schrödinger equation of free electrons in the presence of the electromagnetic field,

$$i\hbar \frac{\partial}{\partial t}\psi(z,t) = (H_0 + H_I)\psi(z,t). \tag{S1}$$

The free-electron Hamiltonian for one-dimensional relativistic dynamics is $H_0 = \varepsilon_0 + v_0(p - p_0) + \frac{(p-p_0)^2}{2\gamma^3 m}$, derived from the Dirac equation in the nonrelativistic approximation when the spin index is ignored. In order to avoid the time jitter bringing a large phase difference, we choose the initial electron kinetic energy $\varepsilon_0 = (\gamma - 1)mc^2 = 10 \text{ keV}$, the initial momentum $p_0 = \gamma m v_0$, and the electron velocity $v_0 = \beta c$ with the speed of light $c$, the relative speed $\beta = 0.1949$ and the Lorentz factor $\gamma = 1/\sqrt{1-\beta^2} = 1.0195$. The pulse has a spectral width $2\sigma_E = 0.85\text{eV}$ (a full-width-half-maximum 1eV). The near-field interaction part is $H_I = -\frac{e}{2\gamma m}(A \cdot p + p \cdot A)$ without gauging. Only the longitudinal component of the near-field vector potential impacts ultrafast electron dynamics in the propagation direction (z). In our case, the transverse field components are ignored.

We take a realistic dielectric membrane suggested in the setup of [1,2], whose longitudinal electric field is given by $E(t) = -\frac{\partial}{\partial t}A(t)$, where the vector potential is $A(t) = A_0 \sin(\omega_L t + \phi_L)$ without the distance-dependent term since we use a thin membrane, with $A_0 = -\frac{E_0}{\omega_L}$, the electric field strength $E_0$, the laser frequency $\omega_L$, and the phase delay $\phi_L$. Here, we choose the mid-infrared laser with central wavelength $\lambda_L = 10.33\,\mu m$. The interaction length is assumed to be the same as the thickness of the membrane $L = 50\,nm$.

2. **FESSI spectral phase reconstruction algorithm**

The spectral interference in Eq. 8 (see Fig.S1(a)) can be measured by an electron spectrometer in our FESSI setup. By transforming the spectrum back into the time domain $\tilde{\tau}$, we obtain

$$\tilde{I}(\tilde{\tau}) = FT\{I(E)\} = D^{dc}(\tilde{\tau}) + D^{-ac}(\tilde{\tau}) + D^{+ac}(\tilde{\tau}), \tag{S2}$$

where the d.c. and a.c. terms are $D^{dc}(\tilde{\tau}) = FT\{|\psi(E)|^2 + |\psi(E - \Delta E)|^2\}$, $D^{\pm ac}(\tilde{\tau}) = FT\{|\psi(E)\psi(E + \Delta E)|\exp\{\mp i\,[\phi_E(E) - \phi_E(E - \Delta E) - E\tau/\hbar]\}\}$. To fetch out the a.c. term, we use a filter in the form of the fourth-order super-Gaussian function, $H(t)$, with a full-width half-maximum $\tau$ and centered at $t = \tau$ (Fig. S1(b)). The filtered signal is then obtained from:

$$D^{filter}(\tilde{\tau}) = H(t - \tau)\tilde{D}(\tilde{\tau}) = D^{+ac}(\tilde{\tau}). \tag{S3}$$

The spectral phase difference is simply the argument of the inverse Fourier transform of $D^{filter}(\tilde{\tau})$ (see Fig. S1(c)), given by

$$\phi_E(E) - \phi_E(E - \Delta E) - E\tau/\hbar = \arg[IFT\{D^{+ac}(\tilde{\tau})\}]. \tag{S4}$$

Note that the extracted phase $\arg[IFT\{D^{+ac}(\tilde{\tau})\}]$ is the phase $\phi_{FESSI}(E)$ in Eq. 2 of the main text. Similarly, by taking a calibration measurement for two replicas with the same time delay, but without the spectral shearing, we can get the calibration phase $-E\tau/\hbar$ in Fig. 2e. Then, by subtracting the calibration term from the extracted phase, and using the concatenation method [3], the spectral phase profile can be obtained (Fig. S1(d)). To this end, both the spectral amplitude and phase are measured, so that we can perform the inverse Fourier transform to the spectral profile (Eq. 1) and reconstruct the full temporal wavefunction.

Concatenation is one of the general methods to reconstruct the phase from the phase difference. By subtracting a calibration phase, we can obtain the spectral phase difference

$$\theta(E) = \phi(E) - \phi(E - \Delta E). \tag{S5}$$

First, we can set the spectral phase at certain position of the energy spectrum equal to zero, so that $\theta(E_0) = -\phi(E_0 - \Delta E)$. Then the spectral phase with new sampling frequency $\Delta E$ is simply:

$$\phi(E_0 + N\Delta E) = \begin{cases} \sum_{n=1}^{N} \theta(E_0 + n\Delta E) & \text{if } N > 0 \\ 0 & \text{if } N = 0 \\ -\sum_{n=N+1}^{0} \theta(E_0 + n\Delta E) & \text{if } N < 0 \end{cases} \quad (S6)$$

Notice that $\theta(E_0 + N\Delta E)$ is already known from Eq. S5, and $E_0$ does not have to be the central energy. By the concatenation method, we can obtain the spectral phase with a sampling frequency equal to the shearing energy.

Figure S2 shows the spectral interference with a larger time jitter of about 0.007% delay time $\tau$. As compared with the time jitter in Fig. 3 in the main text, here we increase the jitter and find that the spectral interference is weaker, but we still find high fidelity F = 98.1%, indicating the robustness of our method.

Figure S3 shows reconstructed free electron wavefunction with phase of primarily 2nd and 3rd orders and some phase oscillations. The high-fidelity value for the comparison between and original and reconstructed phases shows the possibility of detecting an oscillatory spectral phase. Although it is difficult to reconstruct a strong oscillatory spectral phase since it will lead to nonlocal pulse trains and LEM failing to conduct a uniform energy shear, it is still possible to reconstruct a oscillatory spectral phase with a small amplitude (see also Fig. S4(c)).

3. **Analysis of temporal pulse duration of the constructed electron wavefunction**

In the main text, we demonstrate that the LEM device can achieve spectral shearing for an electron pulse under the conditions $\sigma_E \gg \hbar\omega_L$. In fact, net spectral shearing also necessitates a severe constraint on the spectral phases. To understand the effect of spectral phases, we can consider the following simple classical particle acceleration picture: the spectral phase can expand the temporal pulse duration of an electron pulse, and when the expanded width is greater than a quarter of the MIR light optical cycle ($\frac{T}{4}$), there will be no net energy transfer or spectral shearing for the electron pulse, even though its intrinsic pulse width is small. To be more specific, we show that a more accurate condition is given by [4]

$$\sigma_t = \sigma_t(\sigma_E, \phi_2, \phi_3) < \frac{T}{4}, \quad (S7)$$

where the expanded pulse duration $\sigma_\tau$ depends on the intrinsic energy width $\sigma_E$, the second-order spectral phase $\phi_2$ and the third-order spectral phase $\phi_3$. It should be noted that the duration $\sigma_t$ is independent of the constant phase $\phi_0$ and the first order phase $\phi_1$. The temporal wavefunction configuration and its pulse broadening in terms of spectral phases will be demonstrated in the following discussion.

The first order term ($\phi_1$) is solely relevant to the choice of the initial time $t_0$. We assume that the wavefunction is given by

$$\psi(t) = \frac{1}{\sqrt{N}} \int \frac{dE}{\hbar} \psi(E) e^{i\frac{E}{\hbar}t}, \quad (S8)$$

where $N$ is the normalized factor. We consider a first order phase, resulting in the following temporal wavefunction:

$$\tilde{\psi}(t) = \frac{1}{\sqrt{N}} \int \frac{dE}{\hbar} \psi(E) e^{i\phi_1(E-E_0)} e^{i\frac{E}{\hbar}t} = \frac{1}{\sqrt{N}} \int \frac{dE}{\hbar} \psi(E) e^{\frac{iE}{\hbar}(t+\phi_1\hbar)} e^{-i\phi_1 E_0} \quad (S9)$$
$$= \psi(t + \phi_1\hbar) e^{-i\phi_1 E_0}$$

The first order phase provides a time delay $\phi_1\hbar$ and a global phase $e^{-i\phi_1 E_0}$. It only gives the spectral interference signal a constant phase levitation in FESSI, which is readily removed by resetting the phase.

The second-order spectral phase ($\phi_2$) can result in pulse chirping in the time domain. A second-order phase is physically meaningful because a propagating electron can accumulate this phase $\phi_2$ due to its nonrelativistic energy dispersion. We consider the following electron pulse with Gaussian shape and second-order phase:

$$\psi(t) = \left(\frac{\hbar}{(2\pi)^{\frac{3}{2}}\sigma_E}\right)^{\frac{1}{2}} \int \frac{dE}{\hbar} \exp\left\{-\frac{(E-E_0)^2}{4\sigma_E^2}\right\} e^{i\frac{1}{2}\phi_2(E-E_0)^2} e^{i\frac{E}{\hbar}t}$$

$$= \left(\frac{\hbar}{(2\pi)^{\frac{1}{2}}(1/(4\sigma_E) - i\phi_2\sigma_E/2)}\right)^{\frac{1}{2}} \exp\left\{-\frac{t^2}{4\sigma_t^2}\right\} e^{-iat^2}, \quad (S10)$$

Where $\sigma_t^2 = \sigma_{t0}^2 + \left(\frac{\phi_2}{2\sigma_{t0}}\right)^2$ and $a = \frac{\phi_2}{8\left(\sigma_{t0}^4 + \left(\frac{\phi_2}{2}\right)^2\right)}$ [5,6]. In this scenario, the temporal duration can be defined as the root-mean-square (rms) width: $\sigma_t = \sqrt{\langle t^2 - \langle t \rangle^2 \rangle}$, where $\langle \cdot \rangle$ denotes the expectation value and takes the form of $\langle t^n \rangle = \frac{\int t^n |\psi(t)|^2 dt}{\int |\psi(t)|^2 dt}$. $\sigma_{\tau 0}$ is the intrinsic uncertainty of the Gaussian pulse, satisfying $\sigma_{t0} = \frac{\hbar}{2\sigma_E}$. The temporal duration should be less than a quarter of the optical cycle $\sigma_t < \frac{1}{4}T$. Correspondingly, we find three requirements: (i) The intrinsic duration of the pulse should be smaller than half cycle of the light: $\sigma_{t0} < \frac{1}{4}T$, which consistently corresponds to the broad energy width condition $\sigma_E \gg \hbar\omega_L$. (ii) When chirping term is dominant $\left(\frac{\phi_2}{2\sigma_{t0}}\right)^2 \gg \sigma_t^2$, the pulse duration is approximate to $\sigma_\tau \approx \frac{\phi_2}{2\sigma_{t0}}$, and the second-order phase has to satisfy $|\phi_2| < \frac{\sigma_{t0}T}{2}$; (iii) $\sigma_t^2$ has a minimum value $\sigma_t^2 \geq \phi_2$, and thus $\phi_2$ should satisfy $\phi_2 < \frac{1}{64}T^2$.

Fig. S5(a) presents a parameter diagram of temporal duration $\tilde{\sigma}_t$ dependence of $\phi_2$ and electron spectral width $\sigma_E$. The white curve is the contour line for a value of $\sigma_t = \frac{T}{4} = 8.61\text{fs}$, corresponding to a quarter of an optical cycle of MIR laser with a center wavelength of $\lambda_L = 10.33\mu\text{m}$. Thus, there is obviously a lower boundary and an upper boundary for $\sigma_E$. This indicates that $\sigma_E$ cannot be too small, otherwise the coherence length will surpass the optical cycle, leading to the failure of spectral shearing. Furthermore, for a given value of second order phase $\phi_2$, $\sigma_E$ has an upper limit because the large spectral width with fixed $\phi_2$ rapidly broadens the temporal pulse.

It is also important to consider a third-order phase ($\phi_3$) because electron acceleration by a high-voltage bias can create $\phi_3$. Similarly, we begin with Gaussian pulse having a third-order phase:

$$\psi(t) = \left(\frac{\hbar}{(2\pi)^{\frac{3}{2}}\sigma_E}\right)^{\frac{1}{2}} \int \frac{dE}{\hbar} \exp\left\{-\frac{(E-E_0)^2}{4\sigma_E^2}\right\} e^{i\frac{1}{6}\phi_3(E-E_0)^3} e^{i\frac{E}{\hbar}t} \quad \text{(S11)}$$

This temporal wavefunction does not have an analytical expression. However, the pulse duration can be explicitly obtained, which has the form $\sigma_t = \sigma_{\tau 0}\sqrt{1 + \frac{1}{2}\left(\frac{\phi_3}{4\sigma_{t0}^3}\right)^2}$ [5,6]. In addition, we numerically show a temporal duration dependence of the third order phase $\phi_3$ and the spectral width $\sigma_E$, as shown in Fig. S5(b). Similar to the second order phase, the third order phase also has a lower and upper boundary for $\sigma_E$. Therefore, the spectral shearing only operates in the regime within the border.

Besides, we found that the pulse duration due to a combination of second- and third order phases is given by

$$\sigma_t = \sigma_{t0}\sqrt{1 + \left(\frac{\phi_2}{2\sigma_{t0}}\right)^2 + \frac{1}{2}\left(\frac{\phi_3}{4\sigma_{t0}^3}\right)^2} \quad \text{(S12)}$$

This expression is obtained by solving the autocorrelation function of the spectral wavefunction [6]. When we compare the contributions of $\phi_2$ and $\phi_3$, we find that the second-order phase leads to the dominant consequence to the pulse broadening. Fig. S4(a) and Fig. S4(b) show typical examples of the spectral wavefunctions with solely a second-order phase and solely a solely third-order phase, respectively, which can be well reconstructed by FESSI.

Notably, it is worth mentioning that as a spectral interferometry approach our FESSI setup possesses the advantage of monitoring ultrashort attosecond electron pulses. Fig. S5 shows a reconstruction example of an attosecond electron pulse with a spectral width of 8.5 eV (10eV FWHM), accompanied by a second and third order phase dispersion. The corresponding FWHM duration is around 250 as. The comparison of the original and reconstructed phases, as well as the high fidelity, demonstrate that our FESSI have a high capability for attosecond electron pulse reconstruction.

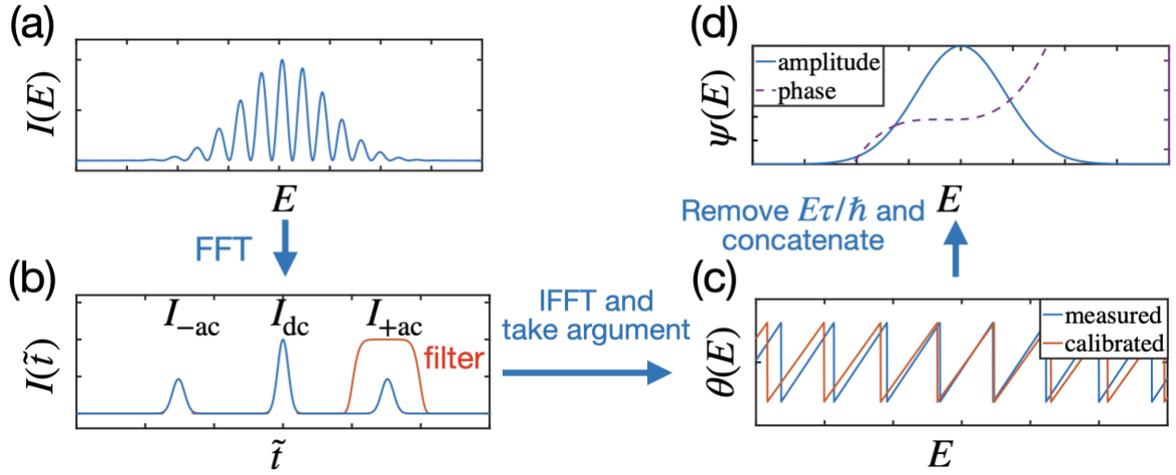

FIG. S1. Flowchart of the reconstruction algorithm for FESSI. (a) Spectral interference of the wavefunction replicas, which is shown in Fig. 1(b) in the main text. (b) Fourier transformation of the spectral shearing interference back to time domain, in order to fetch out the a.c. term. (c) We extract the phase by taking the argument of inverse Fourier transformation of the a.c. term. (d) The reconstructed spectral phase (purple) is found by removing the calibration phase $E\tau/\hbar$ and further concatenating it.

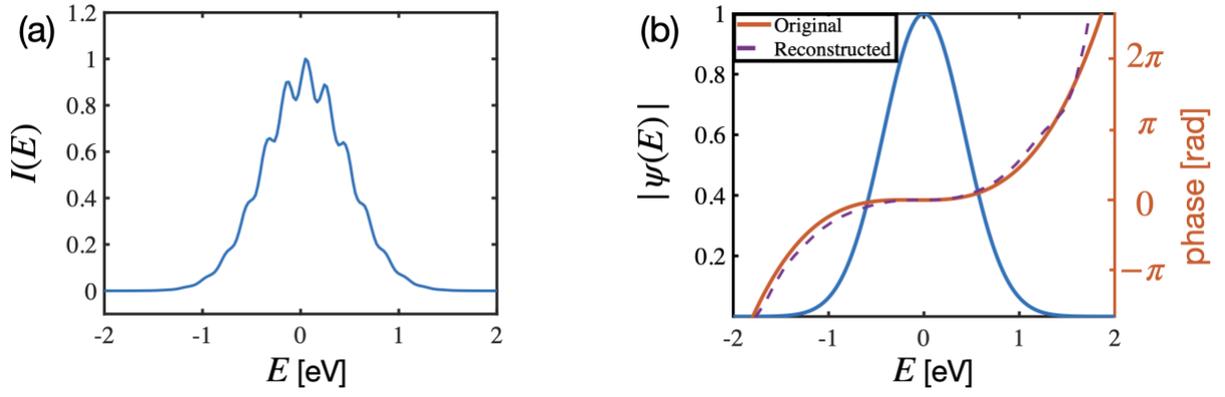

FIG. S2. Reconstructed free electron wavefunction with the consideration of time jitter. (a) shows the simulated interference signal of two replicas of a free electron pulse. The parameters are the same in Fig. 3, except for an increased time jitter around 0.007% of the time delay. (b) shows the spectral amplitude, original phase, and reconstructed phase of the electron wave function. The fidelity value for the reconstruction is $F = 98.1\%$.

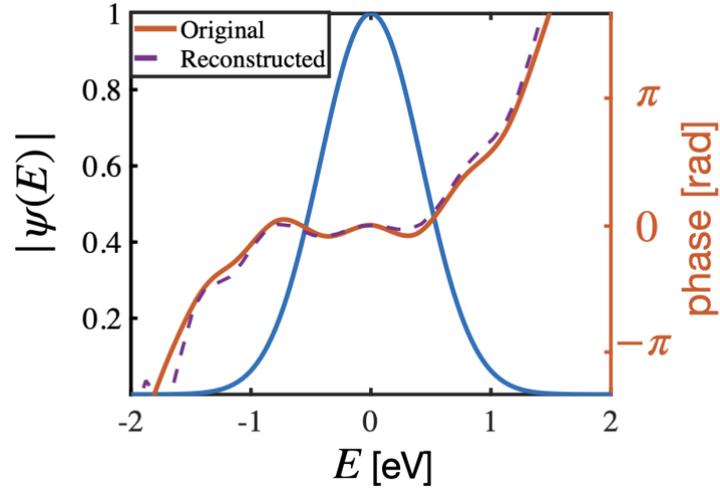

FIG. S3. Original and reconstructed free electron wavefunction with amplitude and phase. It shows an original electron pulse with a spectral width of 0.85 eV (1eV FWHM) and accompanied by a complex phase of primarily 2nd and 3rd orders and some phase oscillations, with the consideration of time jittering around 0.001% of the time delay. The fidelity value for the reconstruction is $F = 97.0\%$.

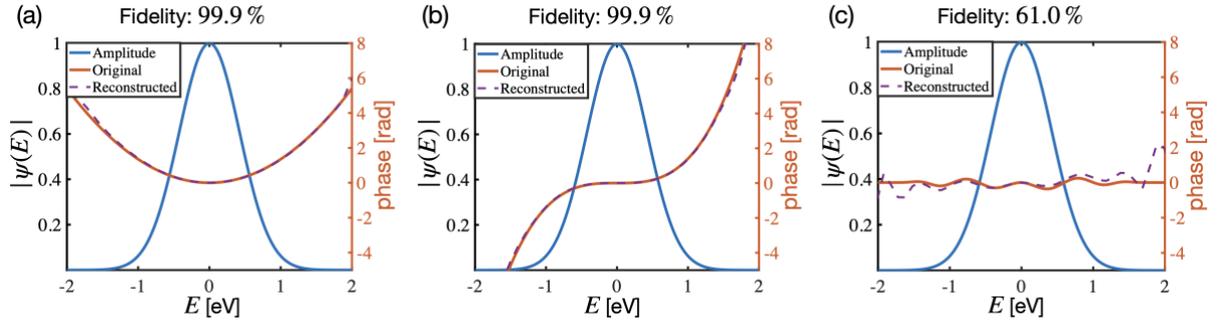

FIG. S4. Original and reconstructed free electron wavefunction with amplitude and phase. It shows an original electron pulse with a spectral width of 0.85 eV (1eV FWHM) and is accompanied by a (a) second phase dispersion $\phi_E(E) = 1.35(E - E_0)^2$ and a (b) third dispersion $\phi_E(E) = 1.40(E - E_0)^3$, and a (c) complicated spectral ringing. Here, we consider the time jittering of roughly 0.001% of the time delay (we repeat 1000 times). The fidelity values for the reconstruction are (a) $F = 99.9\%$, (b) $F = 99.9\%$, and (c) $F = 61.0\%$.

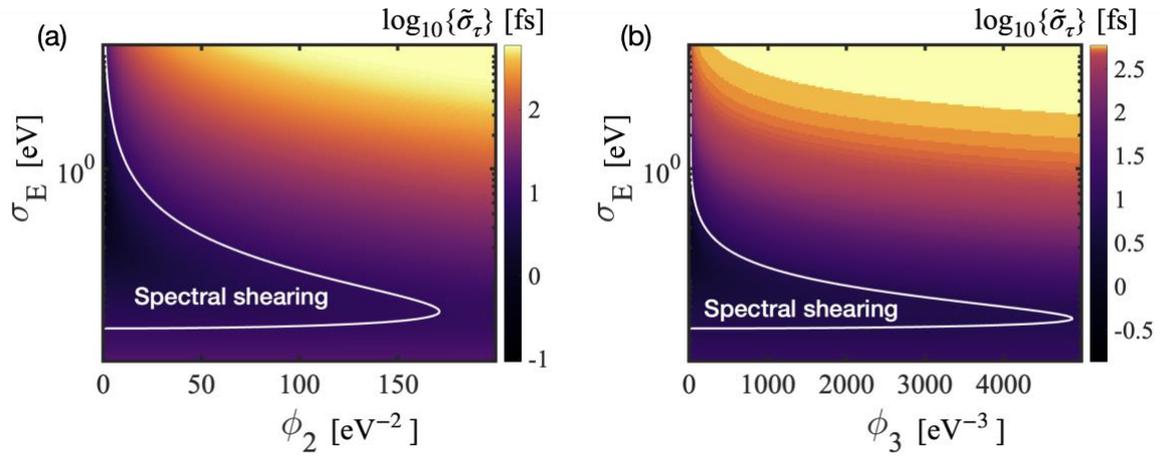

FIG. S5. Parameter diagrams of temporal duration $\tilde{\sigma}_\tau$ dependence of electron spectral width $\sigma_E$ and (a) second order phase $\phi_2$ in Eq. S7, and (b) third order phase $\phi_3$ in Eq. S8. The white contour line has a value of $\sigma_t(\sigma_E, \phi_2, \phi_3) = \frac{T}{4} = 8.61\text{fs}$, corresponding to a quarter of a MIR laser optical cycle with central wavelength $\lambda_L = 10.33\mu\text{m}$.
25

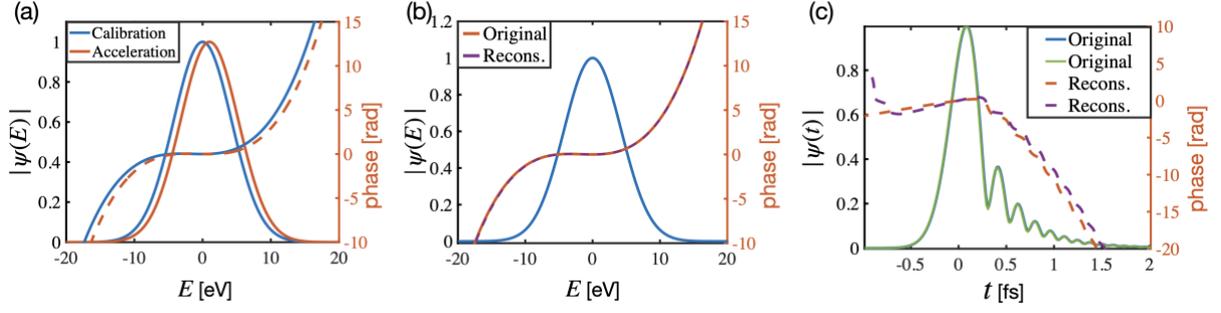

Fig. S6: Original and reconstructed attosecond free electron wavepacket with both amplitude and phase. (a) shows an original electron pulse with a spectral width of 8.5 eV (10eV FWHM) and accompanied by a 2$^{nd}$ and 3$^{rd}$ dispersion ($\phi_E(E) = \phi_2(E - E_0)^2 + \phi_3(E - E_0)^3 = 1.35 \times 10^{-2}(E - E_0)^2 + 2.6 \times 10^{-3}(E - E_0)^3$), and the accelerated one with energy shift $\Delta E = 1$ eV. (b) Our proposed method shows the spectral amplitude and the original and reconstructed phases of the electron wave function. With the consideration of time jitter around 0.001% of the time delay, the fidelity value for the reconstruction is $F = 99.9\%$. (c) Comparison of the original wavefunction and the reconstructed one in time. The time delay considered here is $\tau = 5$fs.